\documentclass[preprint2]{aastex6}
\usepackage{epsfig}
\usepackage{amsfonts,amssymb,amsmath}

\usepackage{caption}
\usepackage{subcaption}

\begin{document}

\title{The Transition Region between Brightest Cluster Galaxies and Intra-Cluster Light in Galaxy Groups and Clusters}
\author{E. Contini$^{1,2}$, H.Z. Chen$^3$ and Q. Gu$^{1,2}$}

\affil{$^1$School of Astronomy and Space Science, Nanjing University, Nanjing 210093, China; {\color{blue} emanuele.contini82@gmail.com, qsgu@nju.edu.cn}}
\affil{$^2$Key Laboratory of Modern Astronomy and Astrophysics (Nanjing University), Ministry of Education, China}
\affil{$^3$Purple Mountain Observatory, No 10 Yuanhua Road, Nanjing 210034, China}

\begin{abstract} 
We take advantage of a state-of-art semi analytic model of galaxy formation, and the model presented in \citet{contini21a}, to investigate the mass distribution of 
Brightest Cluster Galaxies (BCGs) and Intra-Cluster Light (ICL) by addressing two points: (1) the region of transition between a BCG dominated distribution and an 
ICL dominated one, and; (2) the relation between the total BCG+ICL mass and the ICL one alone. We find the transition radius to be independent of both BCG+ICL 
and halo masses, with an average of 60$\pm$40 kpc, in good agreement with previous observational measurements, but given the large scatter, it can be considered as 
a sort of physical separation between the two components only on cluster scale. 
From the analysis of $M_{ICL}-M_{BCG+ICL}$ relation, we build a method able to extract the ICL mass directly from the knowledge of the BCG+ICL one. Given the large 
scatter on low mass systems, such method under/overpredicts the true value of the ICL in a significant way, up to a factor of three in the worst cases. On the other 
hand, for $\log M_{BCG+ICL}>12$ or $\log M_{Halo}>14$, the difference between the true value and the one extracted from the $M_{ICL}-M_{BCG+ICL}$ relation ranges between 
$\pm$30\%. We therefore suggest this relation as a reliable test for observational works aiming to isolate the ICL from the BCG, for systems hosted by haloes on cluster scale. 
\end{abstract}


\section[]{Introduction} 
\label{sec:intro}

The diffuse stellar component in galaxy groups and clusters has only recently being studied deeply, but it was first proposed (\citealt{zwicky37}) and then confirmed by \cite{zwicky51} in the Coma cluster. 
Generally known as Intra-Cluster Light (ICL), or Diffuse Stellar Component (DSC), it is made of stars that are not gravitationally bound to any cluster galaxy and feels only the potential well of 
the dark matter (DM) halo. Although the ICL is spread throughout the whole group/cluster, most of it is usually associated with the Brightest Cluster Galaxy (BCG), which is the galaxy residing at the centre 
of the DM halo, but considerable amounts of ICL are predicted (e.g., \citealt{contini14,contini18}) and found (e.g., \citealt{gonzalez13,presotto14}) to be concentrated around intermediate and massive 
galaxies (for an introductory review on the ICL see \citealt{montes19b} or \citealt{contini21b}).

The nature of the ICL is still under debate, although several physical processes have been suggested in order to explain its formation and evolution. Disprution of dwarf galaxies 
(\citealt{murante07,purcell07,conroy07,martel12,giallongo14,annunziatella16,morishita17,raj20}), tidal stripping of intermediate and massive galaxies 
(\citealt{rudick09,rudick11,contini14,demaio15,demaio18,montes18,jimenez18,jimenez19}), minor and major mergers between centrals and satellites 
(\citealt{monaco06,murante07,gerhard07,contini14,burke15,groenewald17,jimenez18,jimenez19}), pre-processing and accretion (\citealt{willman04,mihos05,rudick06,sommer-larsen06,contini14}), 
and \emph{in situ} star formation (\citealt{puchwein10}) are the mechanisms that have been invoked during the recent years. Considering the evidence (e.g., \citealt{melnick12}) that rules out \emph{in situ} 
star formation and several studies (e.g., \citealt{montes14,contini14,demaio15,montes18,contini19}) that downgraded disruption of dwarfs as a secondary channel, mergers, stellar stripping and pre-processing 
are now believed to be the most important channels for the formation of the ICL.

The ICL is a faint component in galaxy groups and clusters, and one of the most serious difficulties in studying it and its properties, in particular from the observational side, lies in its own definition.
Indeed, in order to isolate the ICL, one should remove every single contamination, and mask the light coming from satellite galaxies. At this point, the only contribution supposedly not ICL should be provided by 
the BCG. The final separation between the BCG and the ICL is still a challenging task (e.g., \citealt{gonzalez13,kravtsov18,montes18,gonzalez21,contini21a,contini21b}). There are several techniques to do that, both 
theoretical and observational, and all of them with pros and cons. Theoretical methods can use information not available to observers, and this eases the study of the ICL from a theoretical point of view. 
Observationally speaking, there are mainly two methods commonly used: (a) the choice of a surface brightness cut under which the light is assumed to be ICL (e.g., \citealt{feldmeier04,zibetti05,presotto14,montes21,furnell21}); (b) the use of functional forms to distribute the total (BCG+ICL) light and consider ICL dominated the region from where the light profile drops, followed by a plateau (\citealt{gonzalez05,seigar07,donzelli11,giallongo14,cooper15,alamo17,durret19,zhang19,kluge21}).

At some distance the ICL dominates over the BCG, and the radius at which it happens is usually defined as \emph{transition radius} (\citealt{zibetti05,contini21a,montes21,gonzalez21}). The transition radius has been
probed both observationally and theoretically, but given the difficulties mentioned aboved and considering that ICL and BCGs overlap in a region that can be more or less extended, it is just a rough estimation of 
the distance where the ICL distribution overcomes the BCG's one. From the observational side, there are a few estimates of the transition radius \citep{zibetti05,seigar07,iodice16,montes21,gonzalez21}. All these 
authors find that the typical transition radius, i.e. where the BCG+ICL profile shows a break, is around 60-80 kpc. These values have been recently confirmed to be plausible also by the model predictions in 
\cite{contini21a}, where we found that the transition radius between the BCG and the ICL ranges from around 15 kpc to over 100 kpc, depending on BCG properties such as its morphology (bulge/disk dominated), on the 
relative amount of ICL with respect to that of the BCG, and on the dynamical state of the cluster.

In this paper, we use the model originally introduced in \cite{contini20} and further developed in \cite{contini21a} to investigate the transition region between the BCG and the ICL. We define the transition radius 
as close as possible to the observational way, i.e. where the ICL starts to dominate the distribution. Moreover, we introduce a BCG+ICL scale radius, defined as the distance from the BCG centre enclosing a given 
percentage of the total BCG+ICL stellar mass, and compare it with the transition radius by considering their relation with both BCG+ICL and halo masses. In the second part of the analysis, we study the relation 
between the BCG+ICL stellar mass and the ICL one alone for a large sample of haloes spanning a wide range of halo mass, as predicted by our model. The two quantities correlate well, especially at large BCG+ICL/halo masses, therefore, via fitting we quantify the typical error one can make in extrapolating the ICL mass from its relation with the total BCG+ICL one.

The layout of the paper is as follows. In Section \ref{sec:methods} we briefly describe how we model BCG+ICL systems and provide a brief description of the two sets of N-body simulations used. In Section 
\ref{sec:results}  we present our analysis, which will be fully discussed in Section \ref{sec:discussion}, and in Section \ref{sec:conclusions} we summarize our main conclusions.  Stellar masses are computed with 
the assumption of a \cite{chabrier03} Initial Mass Function (IMF), and, unless otherwise stated, all units are $h$ corrected. 

\section{Methods}  
\label{sec:methods}
In this Section we briefly describe how we model the bulge, disk and ICL profiles, and introduce the two quantities used in the first part of the analysis: the transition radius, $R_{transition}$, and the 
BCG+ICL scale radius, $R_{BCG+ICL}$. Besides, given that our samples of BCG+ICL systems come from a semi-analytic model of galaxy formation (the detail of the model can be found in \citealt{delucia07,contini14,contini18}) run on merger trees extracted from N-body simulations, we briefly describe the two sets of simulations used. We stress here that, in this work, we use two sets of simulations with different cosmological parameters.

\subsection{BCG and ICL profiles}
\label{sec:profiles}
As mentioned in Section \ref{sec:intro}, the BCG+ICL light distribution is, observationally, derived by profile fitting with the use of functional forms, usually double/triple Sérsic (\citealt{sersic68}) or composite 
profiles. In order to be consistent with the growths of bulge and disk in the semi-analytic model, in \cite{contini21a} we modelled them by assuming a Jaffe profile \citep{jaffe83} for the bulge, and an 
exponential one for the disk. It is worth stressing that BCGs are normally ellipticals, but in some cases they have time to grow a new disk with the absence of major mergers and a strong AGN feedback. In 
our sample, most of them can be classified as early-type galaxies with negligible disks, but as stated above, to be consistent with the growths of bulge and disk, we consider both components in the analysis.

The Jaffe profile reads as follows:
\begin{equation}\label{eqn:jaffe1}
\rho_J (r) = \frac{M_J r_j}{4\pi r^2 (r_j +r)^2}  ,
\end{equation}
where $M_J$ is the total mass of the bulge and $r_J$ the scale length. By integrating within a sphere of radius $r$, we obtain the mass
\begin{equation}\label{eqn:jaffe2}
M_J (r) = \frac{M_J r_j}{r_j +r}  .
\end{equation}

The exponential profile for the disk, instead, has the following form:
\begin{equation}\label{eqn:disk}
M_D (r) = M_{D, tot} \left[1-\left(1+\frac{r}{R_D} \right)e^{-r/R_D} \right] ,
\end{equation}
where $M_{D, tot}$ is the total stellar mass in the disk, and $R_D$ the scale radius of the disk. 

The distribution of the ICL is modelled by assuming a modified version of the NFW profile \citep{navarro97}:
\begin{equation}\label{eqn:nfw}
\rho (r) = \frac{\rho_0}{\frac{r}{R_s}\left(1+\frac{r}{R_s}\right)^2} \, ,
\end{equation}
where $\rho_0$ is the characteristic density of the halo at the time of its collapse and $R_s$ is its scale radius. The concentration of a DM halo is defined as
\begin{equation}\label{eqn:concentration}
c = \frac{R_{200}}{R_s} \, ,
\end{equation}
where $R_{200}$ is the virial radius of the halo.
The modification with respect to an NFW profile described in Equation \ref{eqn:nfw} is the introduction of a free parameter in Equation \ref{eqn:concentration}. We define the ICL concentration as
\begin{equation}\label{eqn:gamma}
c_{ICL} = \gamma \frac{R_{200}}{R_s} \, .
\end{equation}
For further details about the profiles, and in particular about the free parameter $\gamma$, we refer the reader to \cite{contini20,contini21a}. We just stress here that, being the following analysis 
focused at the present time, the value of $\gamma$ used is 3, which means that the ICL is more concentrated than the DM halo in which it resides. As discussed in \cite{contini20}, this is in 
good agreement with recent theoretical works (e.g. \citealt{harris17,contini18,pillepich18}) and observational studies (e.g. \citealt{montes19}). Moreover, we consider stellar haloes surrounding 
the BCG as being part of the ICL. This caveat is important in the context of comparisons with other studies, both observational and theoretical, which might consider stellar haloes to be either part of the 
BCG or even a third component.

In the following analysis we use two radii: (a) a scale radius of BCG+ICL system, $R_{BCG+ICL,X\%}$, which is defined as the radius that contains $X\%$ of the total BCG+ICL stellar mass, and the transition 
radius, $R_{transition}$, defined as the radius where the differential mass in ICL is 90\% of the total (BCG+ICL). Both radii are derived by using the bulge, disk and ICL profiles described above.

\subsection{N-body simulation sets}
\label{sec:simulations}

In this paper we take advantage of two sets of N-body simulations, from which we extracted the mergers trees that are the input of our semi-analytic model. The two sets are different in many ways: one 
comprises 27 high-resolution zoom-in cluster simulations, and the other one is a cosmological simulation with a box size of around 170 Mpc/h. Below the important details of both simulations.

{\bf DIANOGA}. This set of simulations has been described in detail in \citealt{contini12} (but further information can be found in \citealt{bassini20}), while here we just mention the most important features. DIANOGA comprises 27  high-resolution numerical simulations of 
regions around galaxy clusters, carried out assuming the following cosmological parameters (see \citealt{bonafede11,fabjan11}): $\Omega_m=0.24$ for the matter density parameter, $\Omega_{\rm bar}=0.04$ for the contribution of baryons, 
$H_0=72\,{\rm km\,s^{-1}Mpc^{-1}}$ for the present-day Hubble parameter, $n_s=0.96$ for the primordial spectral index, and $\sigma_8=0.8$ for the normalization of the power spectrum. The latter is expressed as the 
r.m.s. fluctuation level at $z=0$, within a top-hat sphere of $8\,h^{-1}$Mpc radius. For all simulations the mass of each DM particle in the high resolution region is $10^8 \, M_{\odot}/h$, and a Plummer-equivalent 
softening length is fixed to $\epsilon=2.3$ kpc/h in physical units at $z<2$, and in comoving units at higher redshift. The data are stored in 93 output times, spanning a redshift range between 60 and 0. Main haloes 
have been identified using a standard friends-of-friends (FOF) algorithm, while the algorithm {\small SUBFIND} \citep{springel01} has been used to decompose each FOF group into a set of disjoint substructures, 
identified as locally overdense regions in the density field of the background halo. Similarly to previous studies, only substructures that contain at least 20 bound particles are considered to be genuine. At $z=0$, 
our sample of halos counts 361 groups and clusters with virial mass larger than $10^{13} \, M_{\odot}/h$ and up to more than $10^{15} \, M_{\odot}/h$.

{\bf NJ167}. NJ167 is a DM-only cosmological simulation of ($166.67 {\rm \, Mpc/h})^3$ box, carried out with $1024^3$ DM particles from redshift $z=127$ to the present time, by using an improved version of the 
TreePM code P-GADGET-3 \citep{springel05}. It can be considered a medium/high-resolution simulation since the mass of each particle is $3.7\cdot 10^8 M_{\odot}/h$ and the softening length $\epsilon=4$ kpc/h. 
The data have been stored in 128 discrete snapshots and the following Planck 2018 \citep{planck20} cosmology has been used: $\Omega_m=0.31$ for the total matter density, $\Omega_{\Lambda}=0.69$ for the cosmological 
constant, $n_s=0.97$ for the primordial spectral index, $\sigma_8=0.81$ for the power spectrum normalization, and $h=0.68$ for the normalized Hubble parameter. Also for this simulation we used {\small SUBFIND} to 
isolate subhaloes that could retain at least 20 bound particles. Our final sample of haloes at $z=0$ counts 1450 groups and clusters with virial mass larger than $10^{13} \, M_{\odot}/h$ and up to almost
$10^{15} \, M_{\odot}/h$.

\section{Results}
\label{sec:results}

The following analysis is divided in two parts: in the first one we analyze the correlation between the two radii defined in Section \ref{sec:profiles}, while in the second part we study the relations between 
the ICL mass with BCG+ICL/halo masses and try to build an analytic method to extrapolate the ICL mass by knowing the total BCG+ICL, without any assumption on their distributions. Before going to the detail of the 
analysis, it is worth noting that one of the two radii, $R_{BCG+ICL}$ scale, is directly observable since, by definition, it is the radius at which the whole BCG+ICL system encloses a given percentage of the total
luminosity/mass. The transition radius, on the contrary, is not directly observable unless a profile fitting in order to separate the two components is perfomed.

\subsection{Transition-Scale Radii Correlation}
\label{sec:trans_scale}

\begin{figure*} 
\begin{center}
\begin{tabular}{ccc}
\includegraphics[scale=.42]{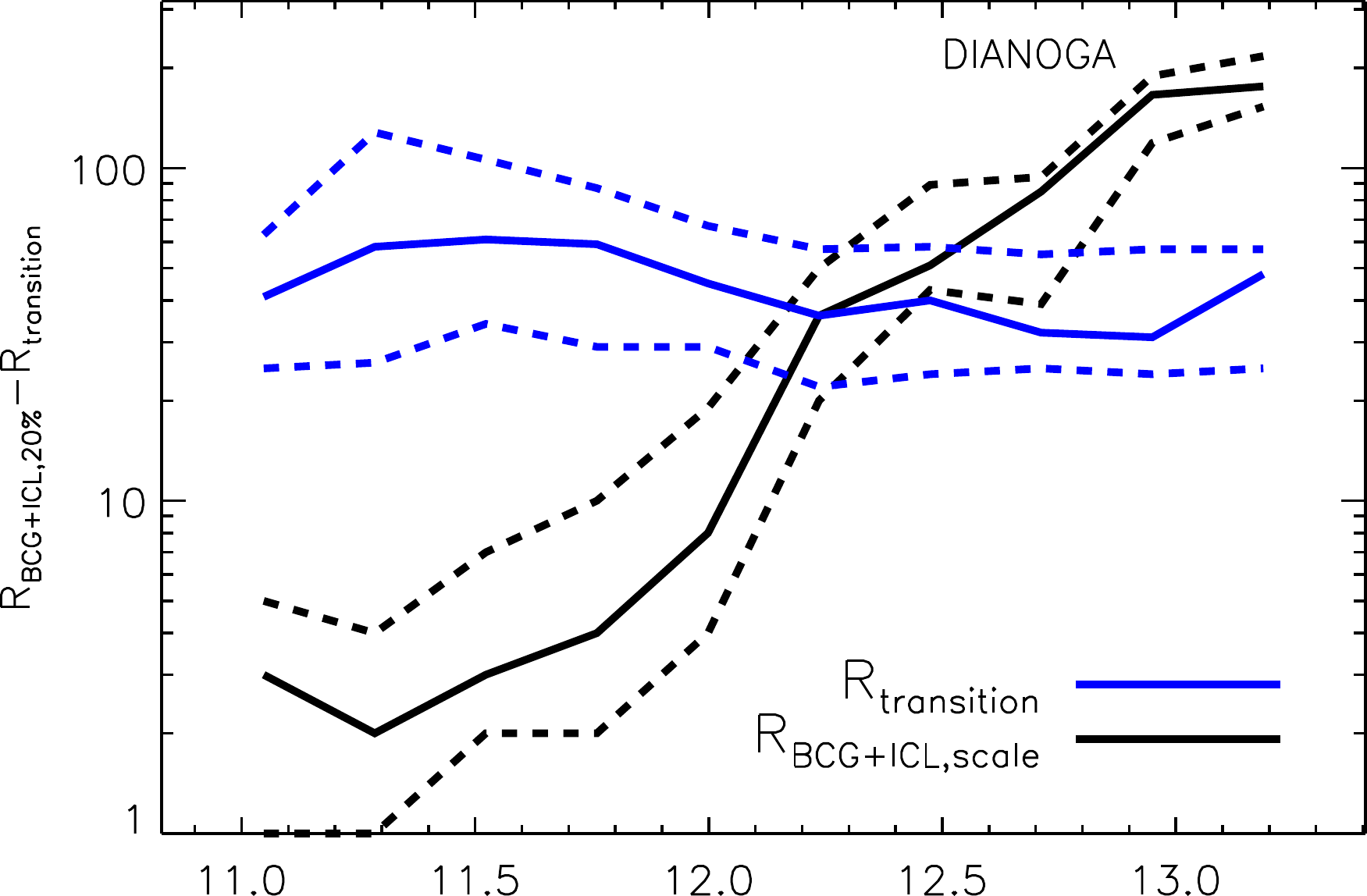} &
\includegraphics[scale=.42]{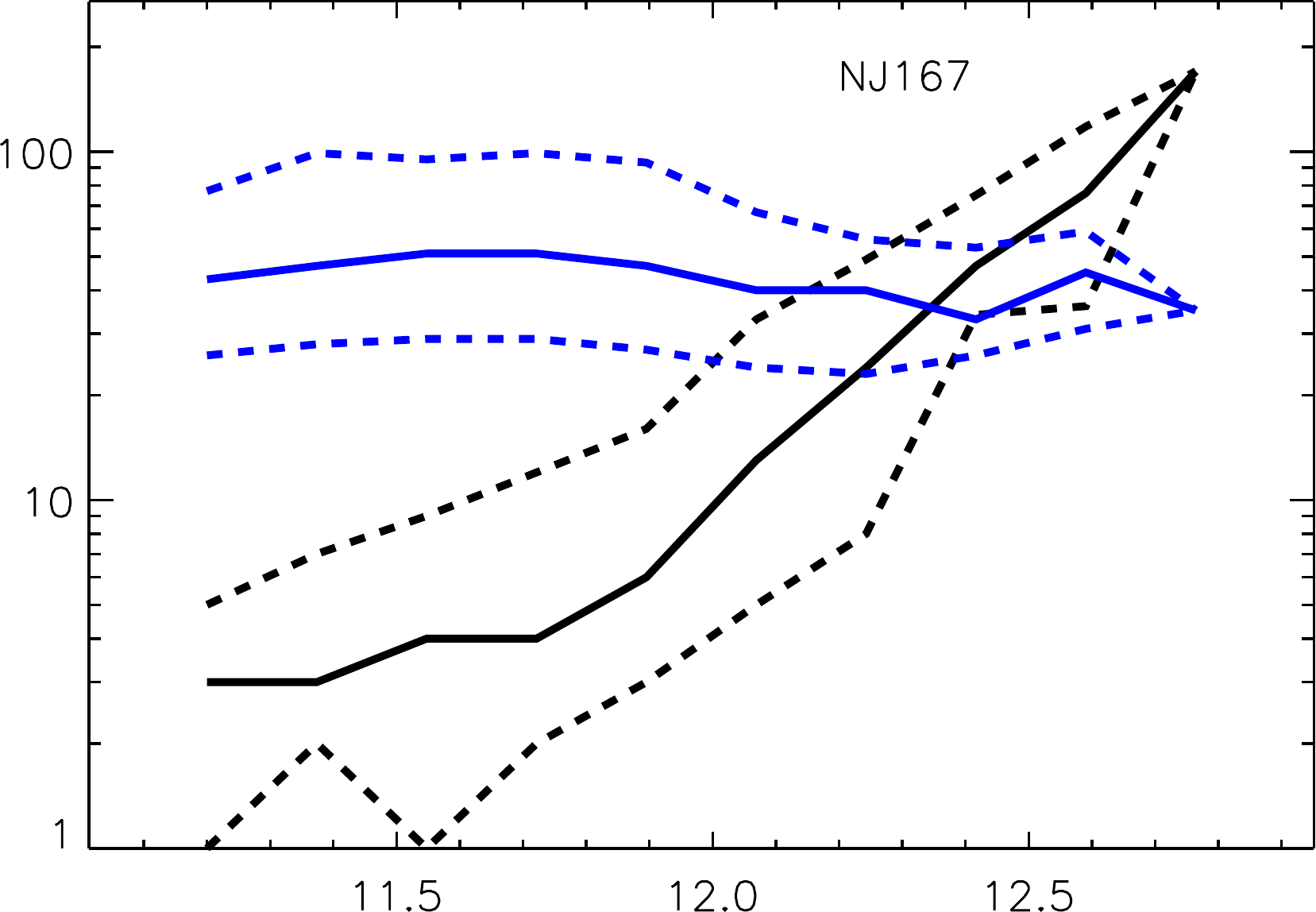} \\
\includegraphics[scale=.42]{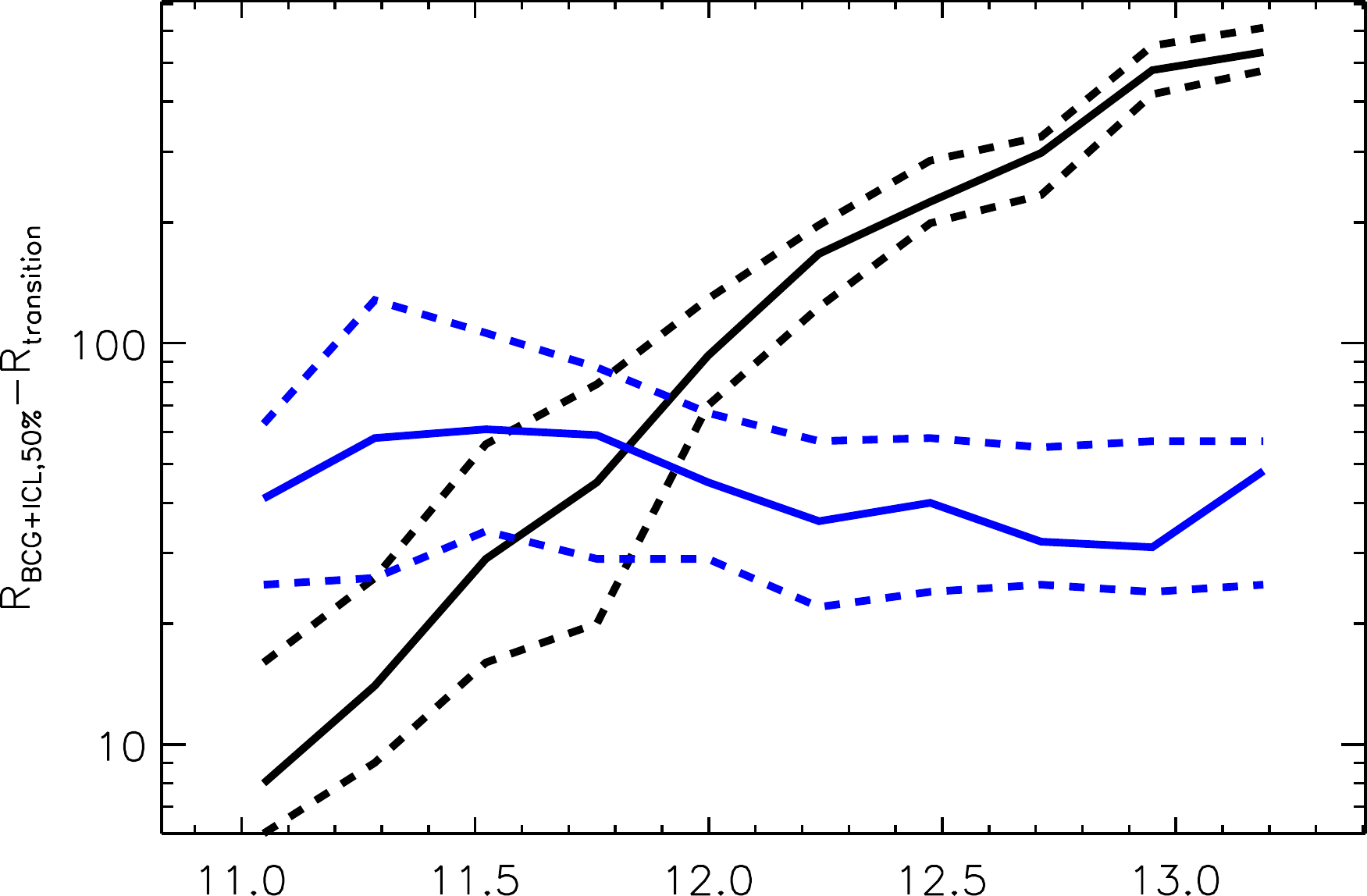} &
\includegraphics[scale=.42]{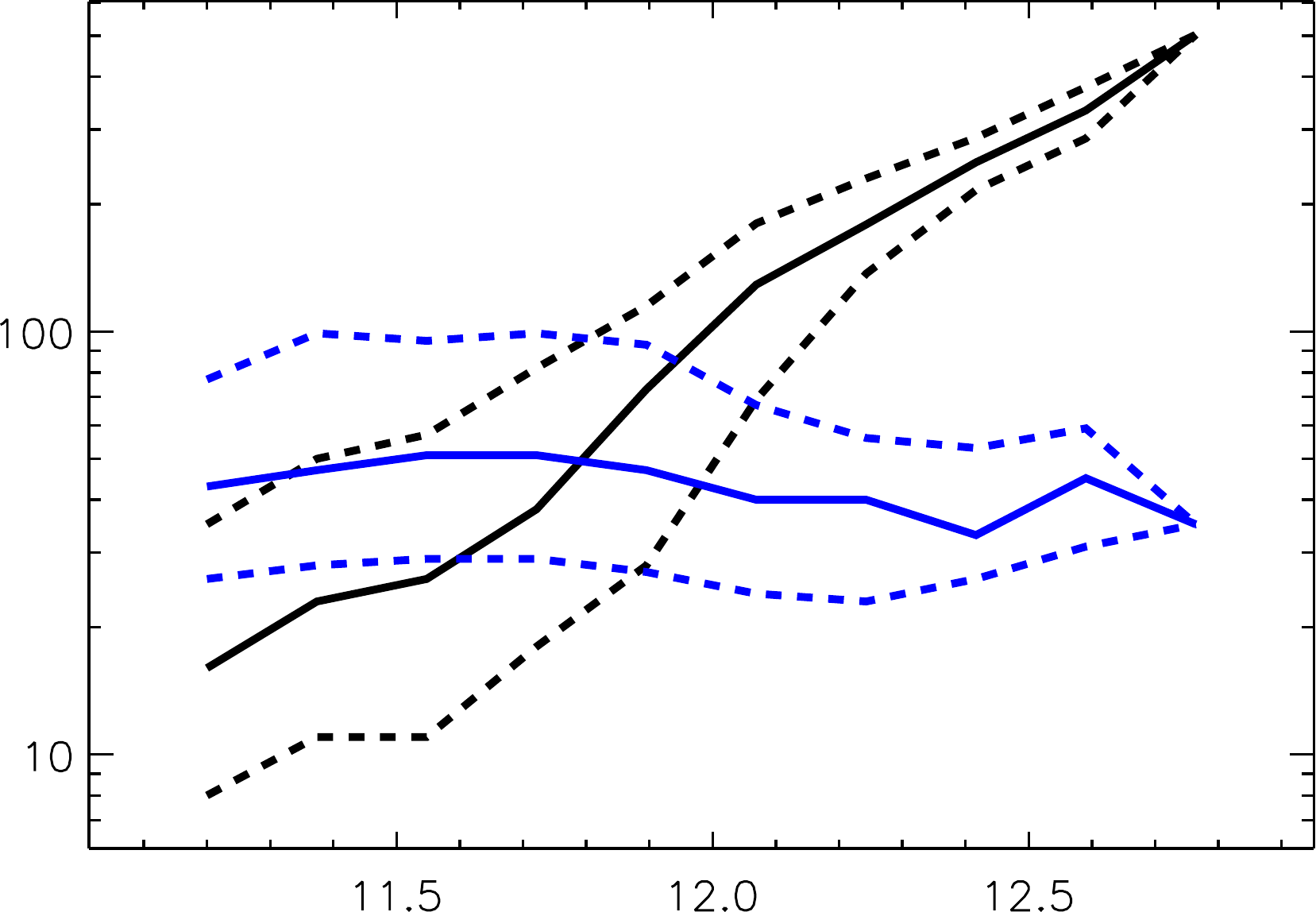} \\
\includegraphics[scale=.42]{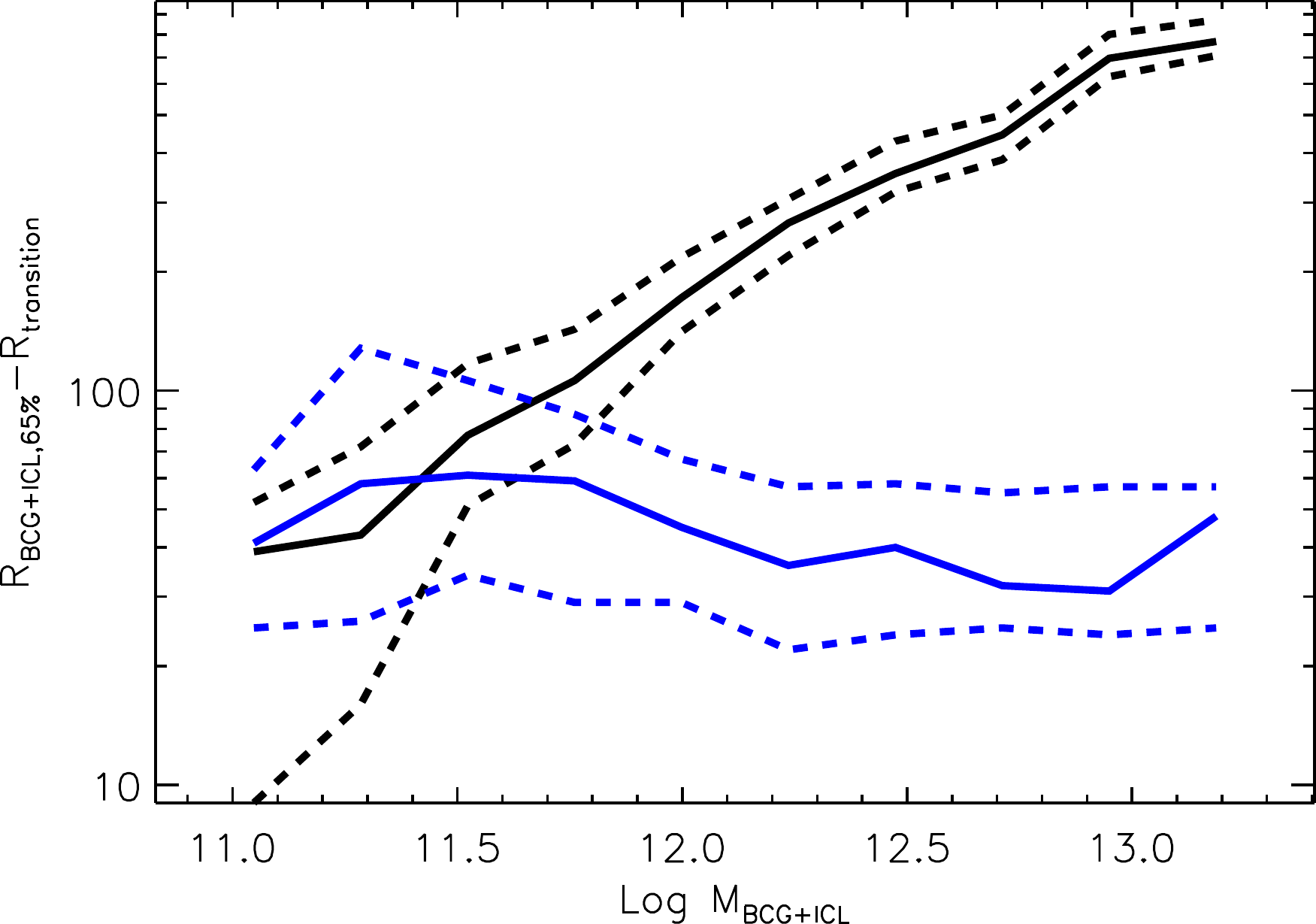} &
\includegraphics[scale=.42]{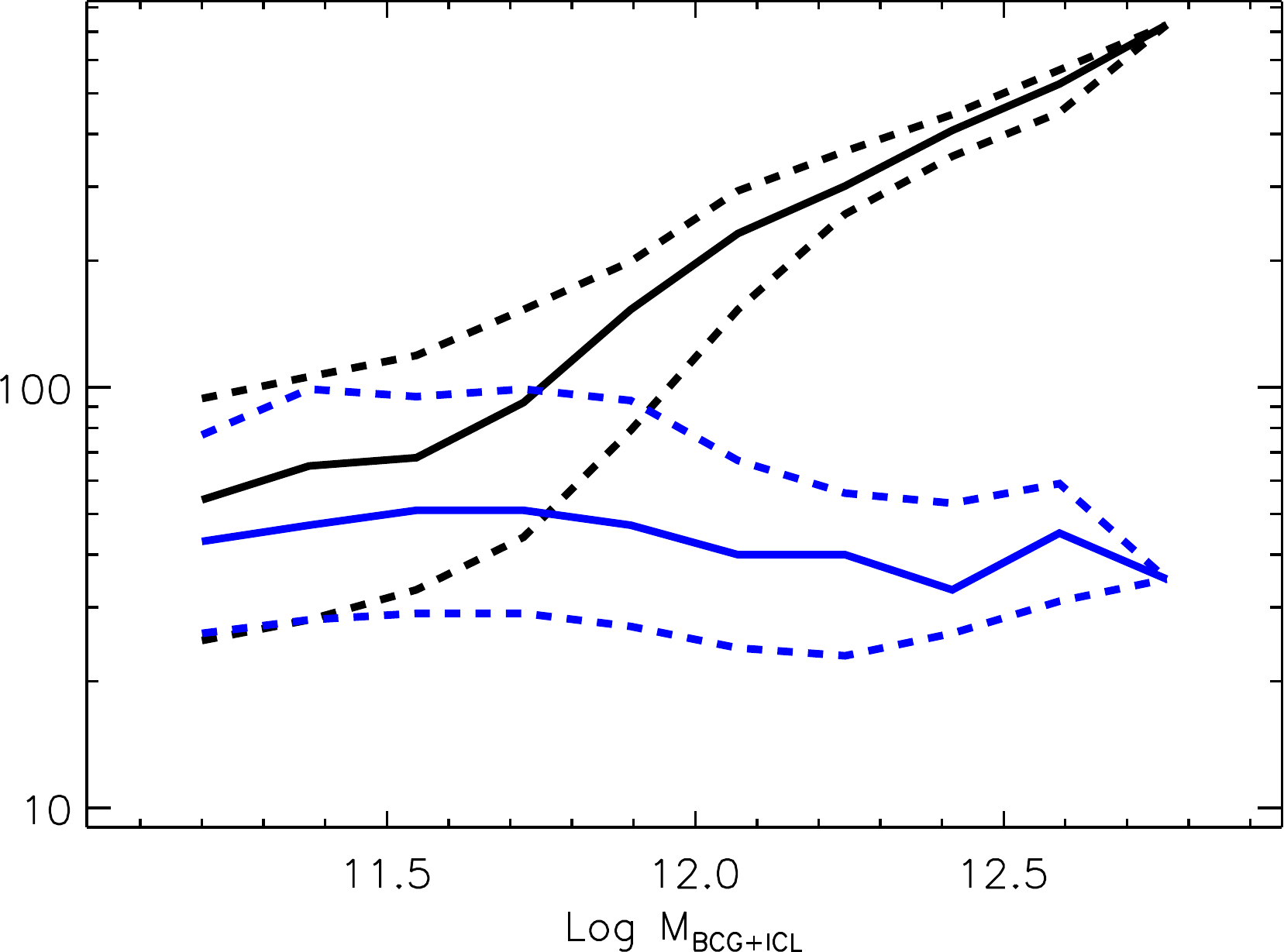} \\
\end{tabular}
\caption{Transition radius $R_{transition}$ (blue lines) and BCG+ICL scale radius $R_{BCG+ICL, X\%}$ (black lines) as a function of BCG+ICL mass, for the two sets of simulations (different columns), and different 
percentages of BCG+ICL mass in the definition of $R_{BCG+ICL, X\%}$ (different rows). Solid lines represent the median of the distributions, and dashed lines represent the 16$^{th}$ and 84$^{th}$ percentiles. 
Regardless the percentage chosen to define $R_{transition}$ (see text for more details), it is independent of the BCG+ICL mass. On the contrary, $R_{BCG+ICL, X\%}$ increases with BCG+ICL stellar mass. Increasing 
the percentage when defining $R_{BCG+ICL}$ leads to sample $R_{transition}$ of decreasing BCG+ICL stellar masses.}
\label{fig:rtrans_bimass}
\end{center}
\end{figure*}

\begin{figure*} 
\begin{center}
\begin{tabular}{ccc}
\includegraphics[scale=.42]{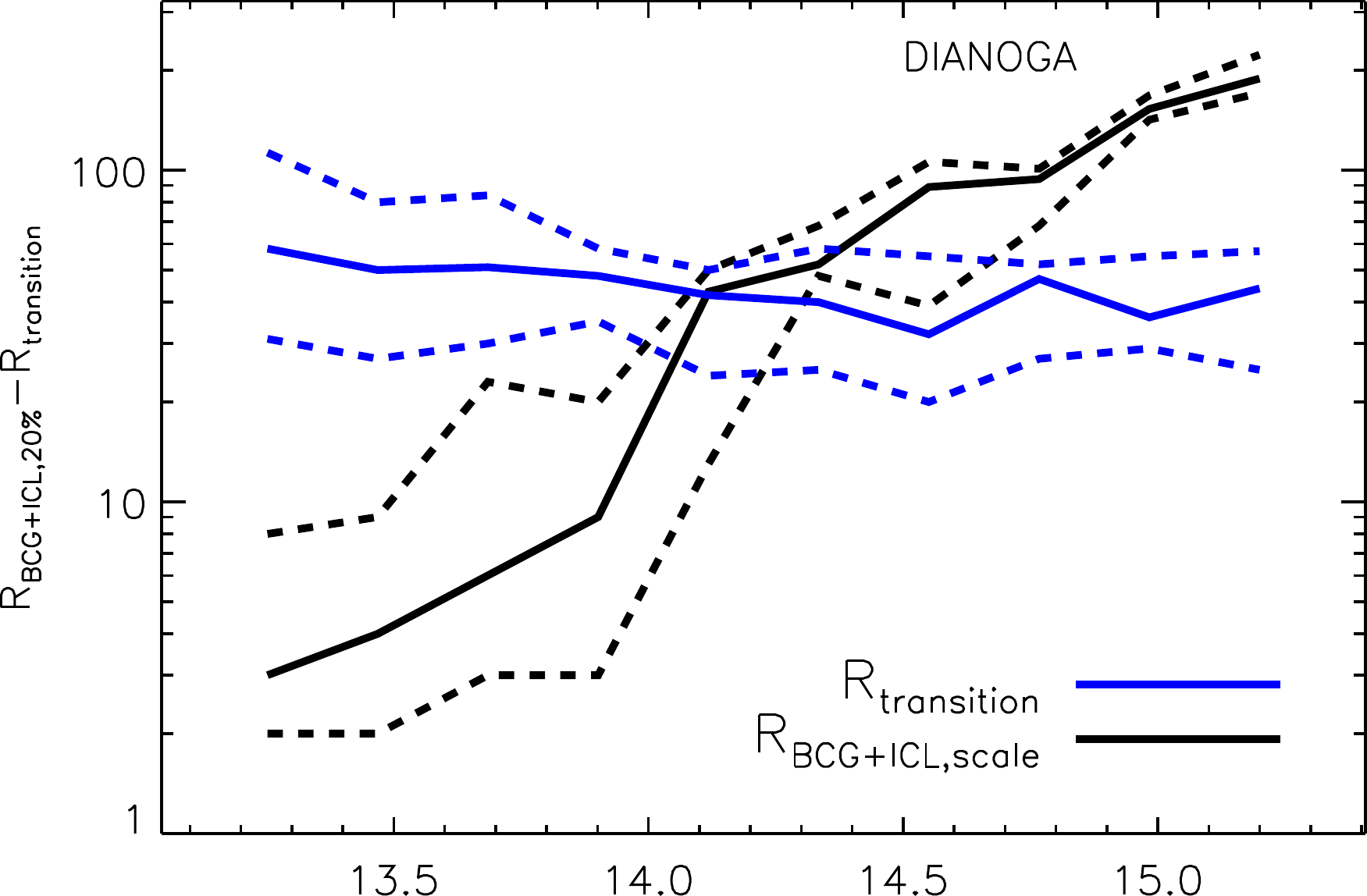} &
\includegraphics[scale=.42]{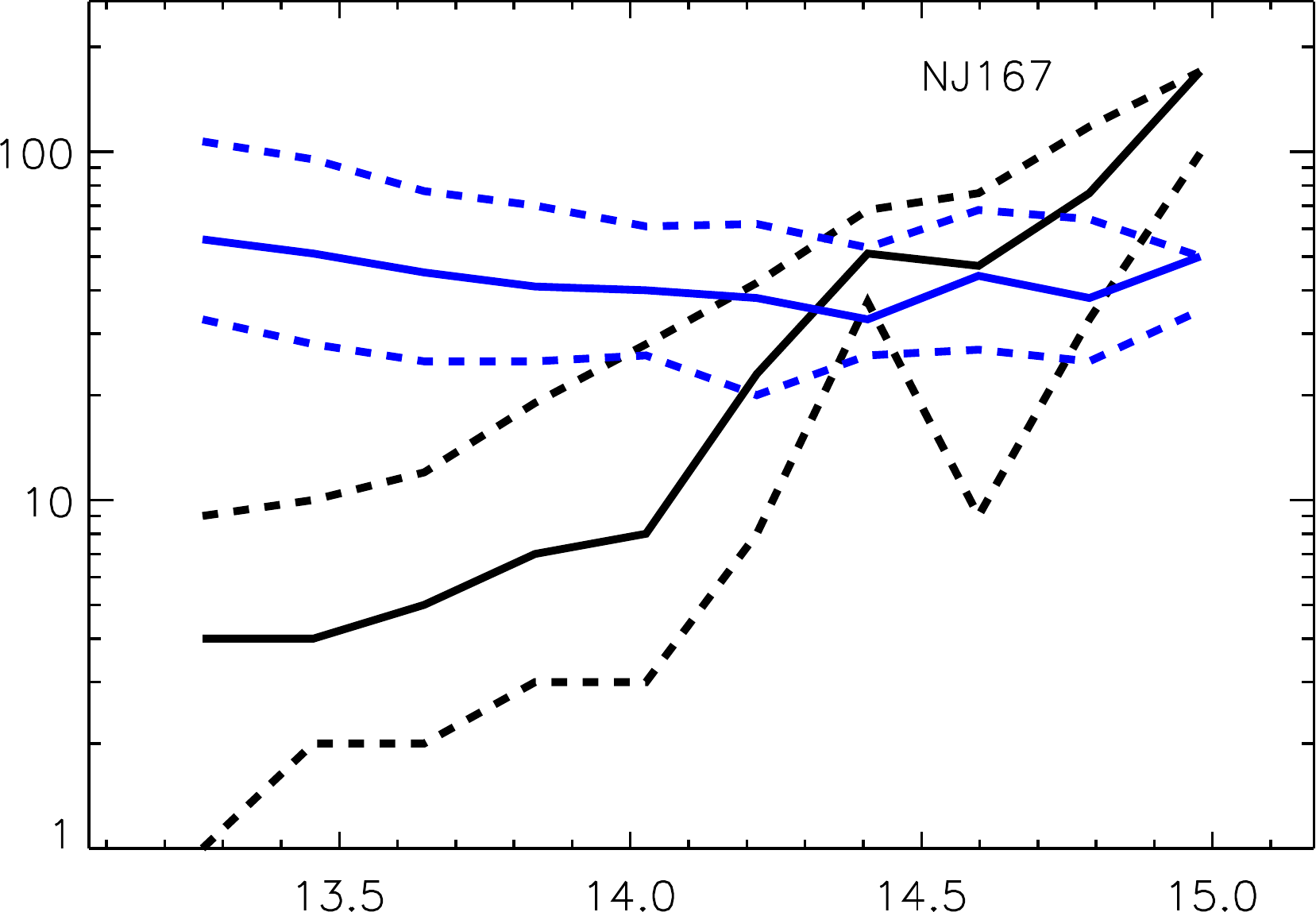} \\
\includegraphics[scale=.42]{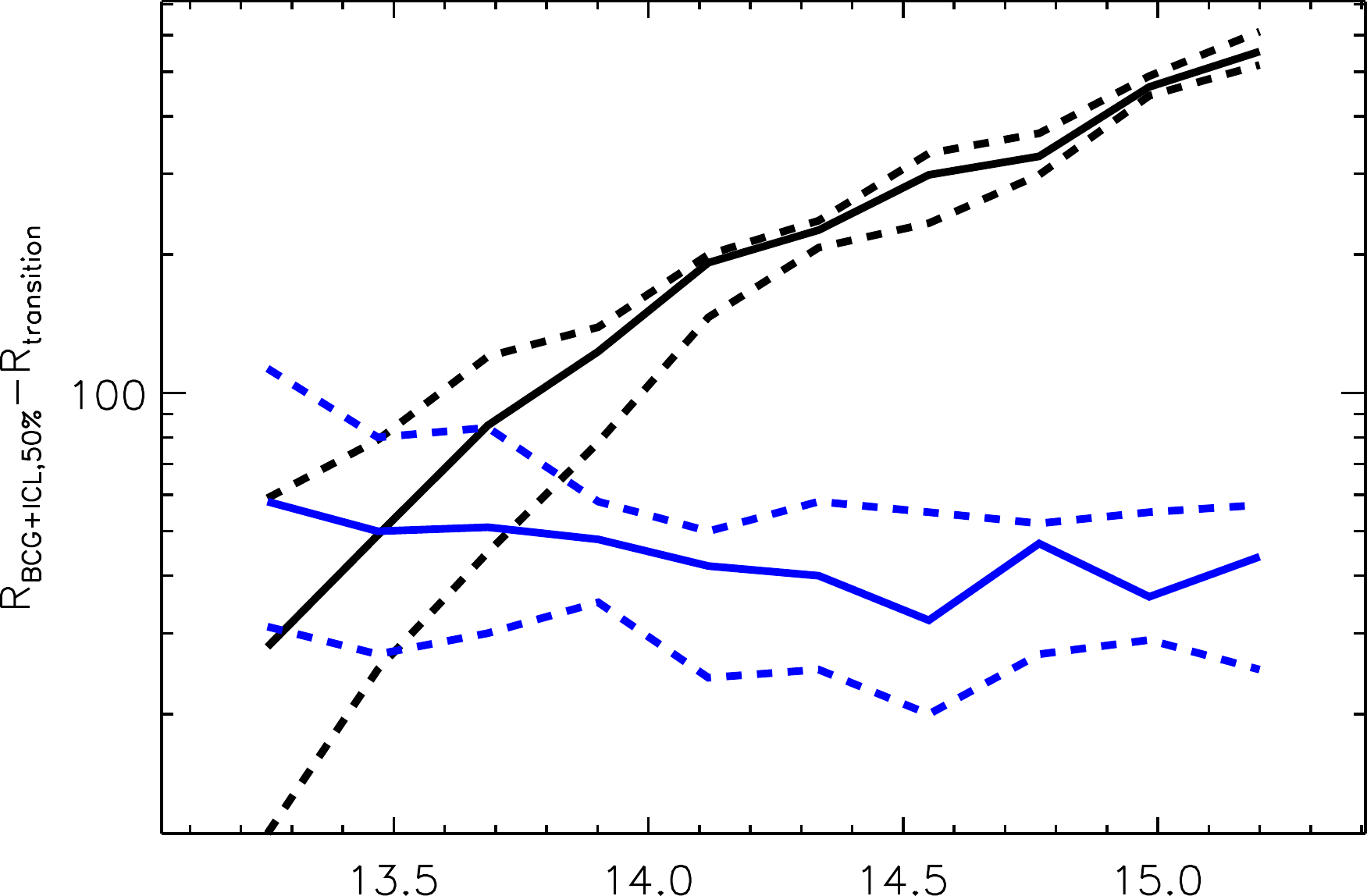} &
\includegraphics[scale=.42]{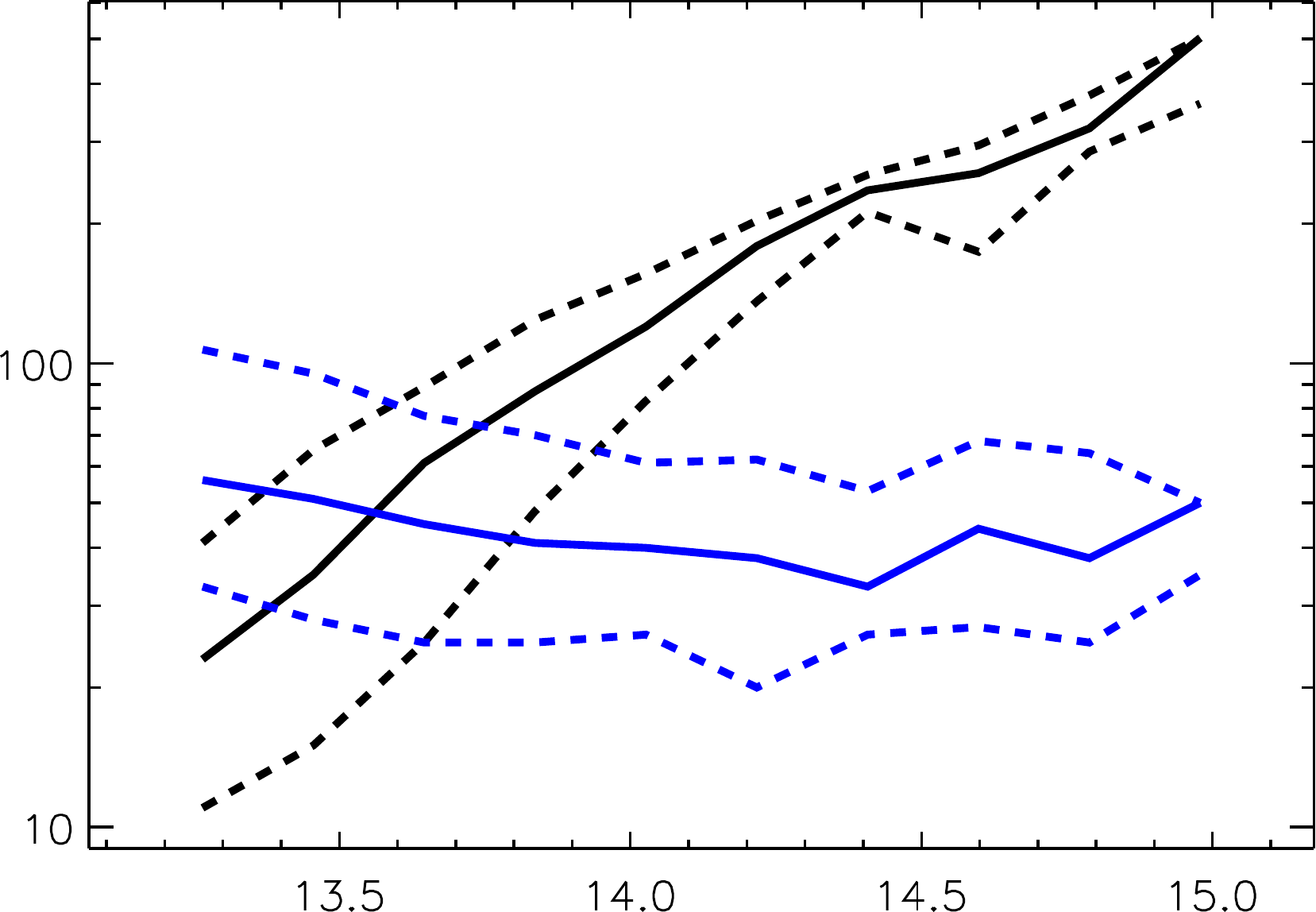} \\
\includegraphics[scale=.42]{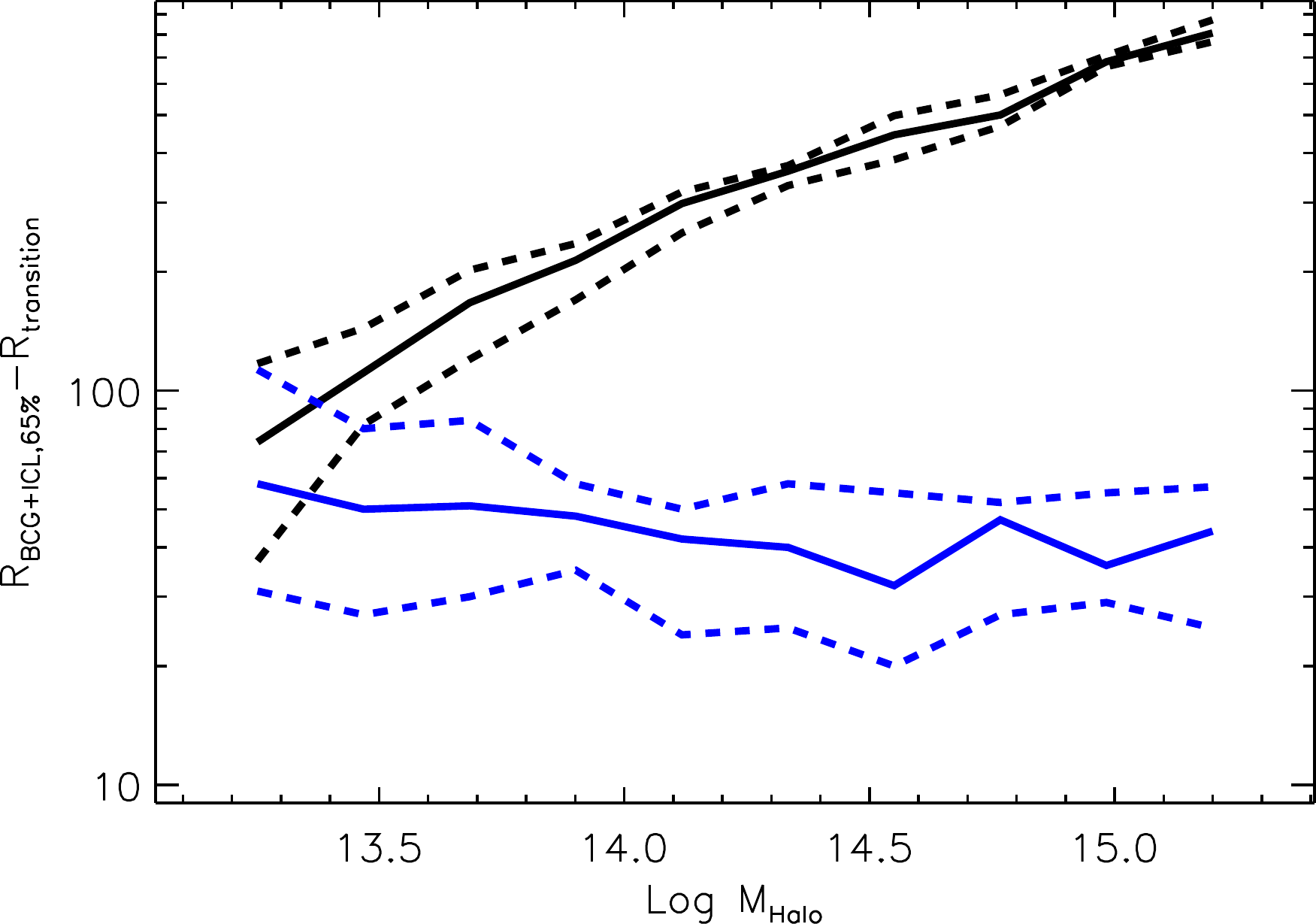} &
\includegraphics[scale=.42]{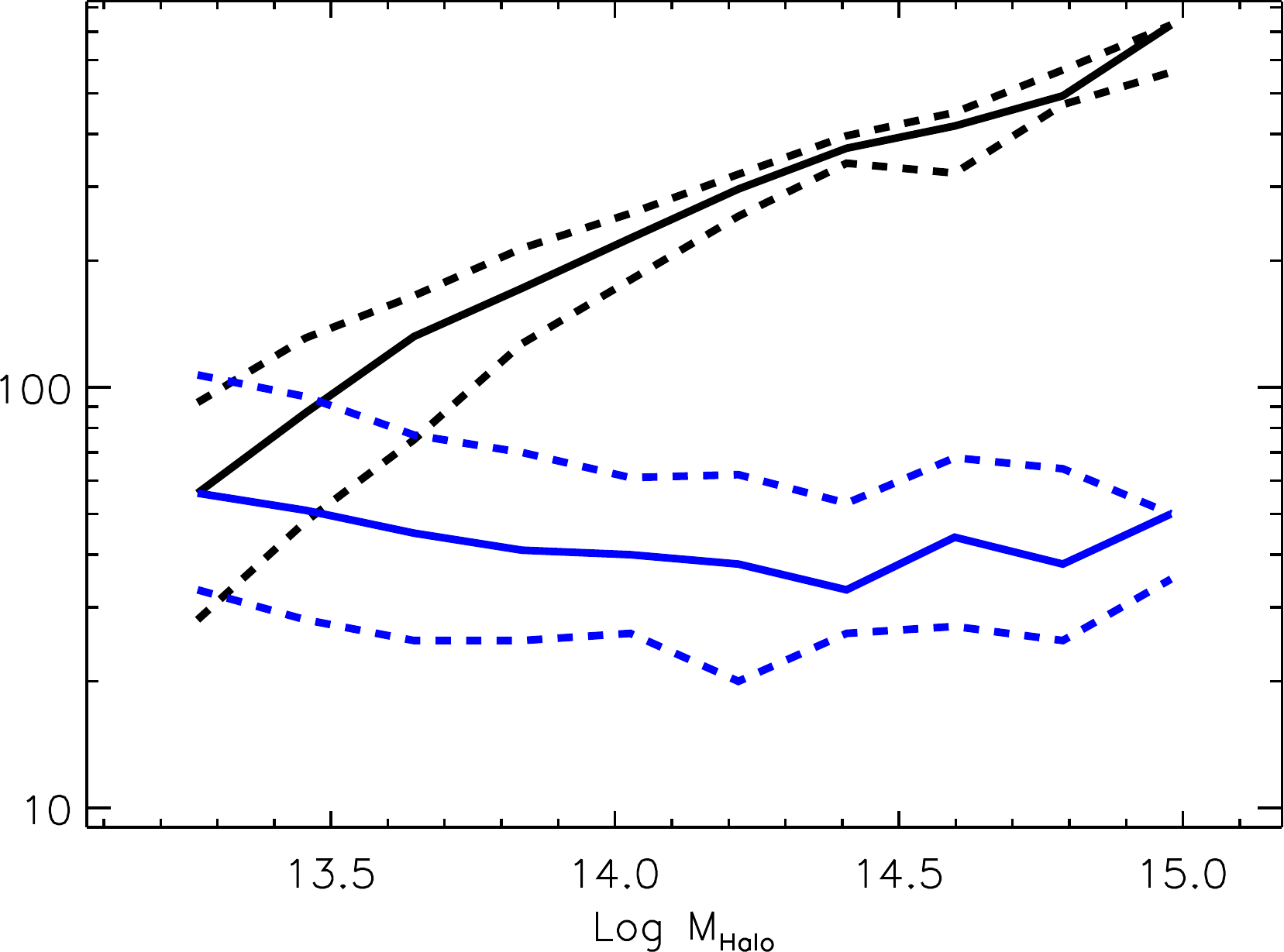} \\
\end{tabular}
\caption{Same information as in Figure \ref{fig:rtrans_bimass}, but as a function of halo mass. Solid lines represent the median of the distributions, and dashed lines represent the 16$^{th}$ and 84$^{th}$ 
percentiles. Also in this case, $R_{transition}$  is independent of the halo mass. Similarly to Figure \ref{fig:rtrans_bimass}, the trend of the percentage used in the definition of $R_{BCG+ICL, X\%}$ with mass 
still holds, in the sense that small percentages sample the transition radius of BCG+ICL systems in high mass haloes, while larger percentages sample those in low mass haloes.}
\label{fig:rtrans_halomass}
\end{center}
\end{figure*}

\begin{table}[hbt!]
\caption{Mean transition radius (and scatter), $R_{transition,X\%}\pm \sigma$, defined as the radius in kpc at which the differential contribution of the ICL is a given percentage of the total (BCG+ICL),
for the two sets of simulations.}
\begin{center}
\begin{tabular}{llllll}
\hline
Simulation & 50\% & 60\%  & 70\% & 80\% & 90\% \\
\hline
\footnotesize{DIANOGA}    & $19\pm 8$   & $24\pm 10$ & $29\pm 13$ & $38\pm 19$ & $58\pm 36$  \\
NJ167                     & $26\pm 16$  & $30\pm 18$ & $35\pm 21$ & $43\pm 26$ & $60\pm 43$  \\       
\hline
\end{tabular}
\end{center}
\label{tab:tab1}
\end{table}

The goal of this section is to link an observable radius, $R_{BCG+ICL}$ scale, with the transition radius, i.e. the radius at which the ICL dominates the total distribution (BCG+ICL), by studying their 
relation with other two \emph{observables}, the total BCG+ICL and halo masses. 

In Figure \ref{fig:rtrans_bimass} we plot the two radii as a function of the total BCG+ICL mass, for different percentages used in the definition of the BCG+ICL scale radius and separately for the two samples 
used (different columns). It appears clear, at first glance, that there is not much difference between the two sets of simulations, except for some scatters. What really appears interesting is that, while the 
transition radius (blue lines) is independent of the BCG+ICL mass, the scale radius (black lines), independently of the percentage chosen to define it, shows a clear trend by increasing from low to high mass BCG+ICL 
systems. Moreover, both radii show a considerable scatter at all stellar masses. Another important feature is that increasing the percentage when defining $R_{BCG+ICL}$ leads to sample $R_{transition}$ at decreasing 
BCG+ICL stellar masses. Indeed, the BCG+ICL scale radius (small percentage) in the top panels samples the transition radius of intermediate-high mass systems ($12.0<\log M_{BCG+ICL}<12.6$), while high percentages, 
such as 65\% used in the bottom panels, sample the transition radius of low mass systems ($11.0<\log M_{BCG+ICL}<11.5$). 

Similar results and trends are found in Figure \ref{fig:rtrans_halomass}, which shows $R_{BCG+ICL}$ and $R_{transition}$ as a function of halo mass. Again, $R_{transition}$  is independent of the halo mass, 
and the trend of the percentage used in the definition of $R_{BCG+ICL, X\%}$ with mass still holds, in the sense that small percentages sample the transition radius of BCG+ICL systems in high mass haloes, while 
larger percentages sample those in low mass haloes. Given the fact that the transition radius is independent of both BCG+ICL and halo masses, we have calculated the typical ranges of $R_{transition}$ for different 
percentages used in its definition, and report them in Table \ref{tab:tab1} separately for DIANOGA and NJ167 simulations. Regardless the percentage adopted, the ranges are all considerably wide, and the mean value 
clearly increases with increasing percentage. When the mass of the BCG and ICL are the same, i.e. 50\% in the definition of $R_{transition}$, the region of transition can be very close to the halo centre. Indeed, the 
mean value for DIANOGA (NJ167) is $19\pm 8$ ($26\pm 16$), which means that in some cases the transition radius can be as small as around 10 kpc. On the other hand, in the extreme case where we consider as transition 
the distance at which the ICL accounts for 90\% of the total (differential) BCG+ICL mass (the case shown in the plots), the transition radius can be as large as 100 kpc, with an average value of around 60 kpc. These results are consistent with what we found in \cite{contini21a} and with several observational measurements \citep{zibetti05,seigar07,iodice16,montes21,gonzalez21}. We will discuss this point in more detail in Section 
\ref{sec:discussion}.

\subsection{ICL-BCG+ICL Correlation}
\label{sec:icl_bcgicl}

\begin{figure}[hbt!]
\includegraphics[scale=0.47]{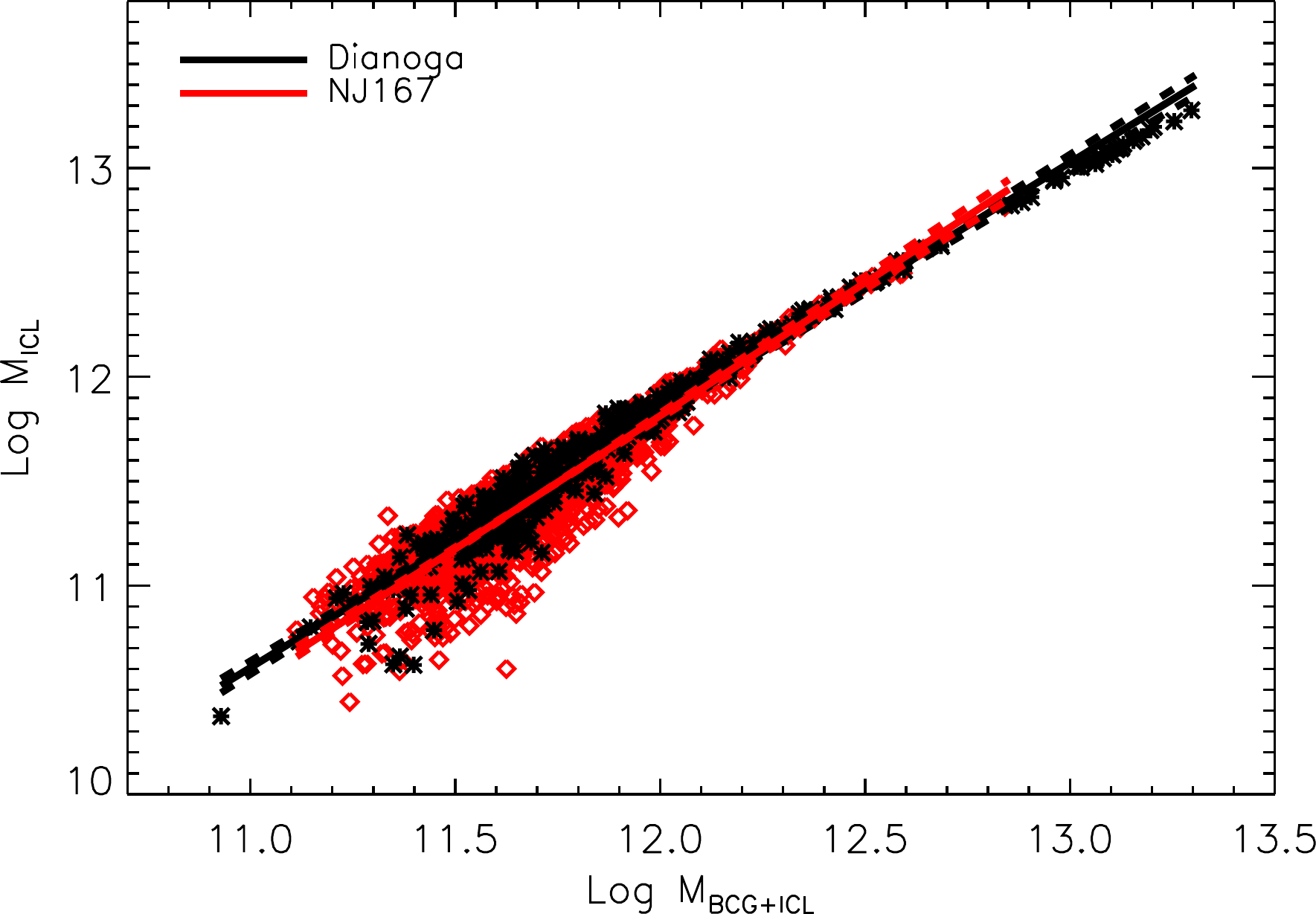}
\centering
\caption{Stellar mass in the ICL versus stellar mass in the BCG+ICL, for Dianoga (black stars and lines) and NJ167 (red squares and lines) simulations. Black and red lines represent the linear fits (including 
$\pm 1\sigma$ scatter) done separately for the two sets of simulations and together, of which details are reported in Table \ref{tab:tab2}.}
\label{fig:micl_mbcgicl}
\end{figure}

\begin{figure}[hbt!]
\includegraphics[scale=0.47]{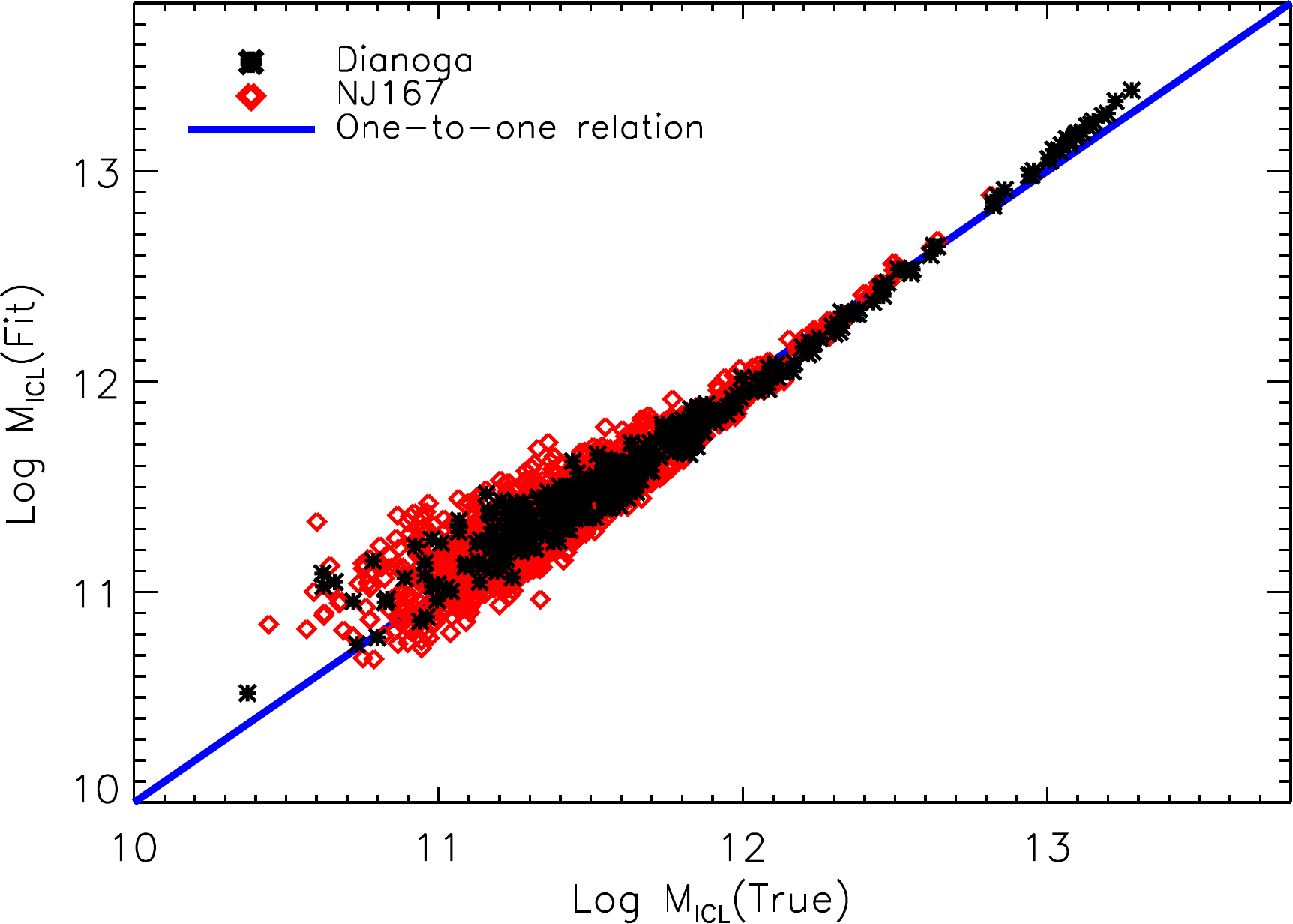}
\centering
\caption{Stellar mass in ICL derived by using the linear fits reported in Table \ref{tab:tab2} and the true value of the ICL mass, for Dianoga (black stars) and NJ167 (red squares). The blue line represents 
a ono-to-one relation between the two quantities. For systems with a true value of ICL mass larger than $\log M_{ICL} \sim 11.6-11.7$, the fits in Table \ref{tab:tab2} provide a reasonable estimate of it, while 
for smaller ICL masses the scatter is quite larger with a trend of overestimation.}
\label{fig:micl_fit-true}
\end{figure}

\begin{table}
\caption{Slopes, intercepts (with $1\sigma$ scatters) and reduced $\chi^2$ of the $M_{ICL}-M_{BCG+ICL}$ relation (Figure \ref{fig:micl_mbcgicl}) for DIANOGA, NJ167 and the two together. Data points have 
been fitted with the linear fit $\log M_{ICL}=\alpha \log M_{BCG+ICL} +\beta$.}
\begin{center}
\begin{tabular}{llllll}
\hline
Simulation & $\alpha$ $\pm 1\sigma$ & $\beta$ $\pm 1\sigma$ & $\chi^2_{\nu}$\\
\hline
DIANOGA      & 1.210 $\pm$ 0.012 & -2.708 $\pm$ 0.134 & 0.010  \\
NJ167        & 1.276 $\pm$ 0.014 & -3.493 $\pm$ 0.158 & 0.016  \\
Together     & 1.251 $\pm$ 0.009 & -3.201 $\pm$ 0.110 & 0.015  \\        
\hline
\end{tabular}
\end{center}
\label{tab:tab2}
\end{table}

\begin{figure*} 
\begin{center}
\begin{tabular}{cc}
\includegraphics[scale=.45]{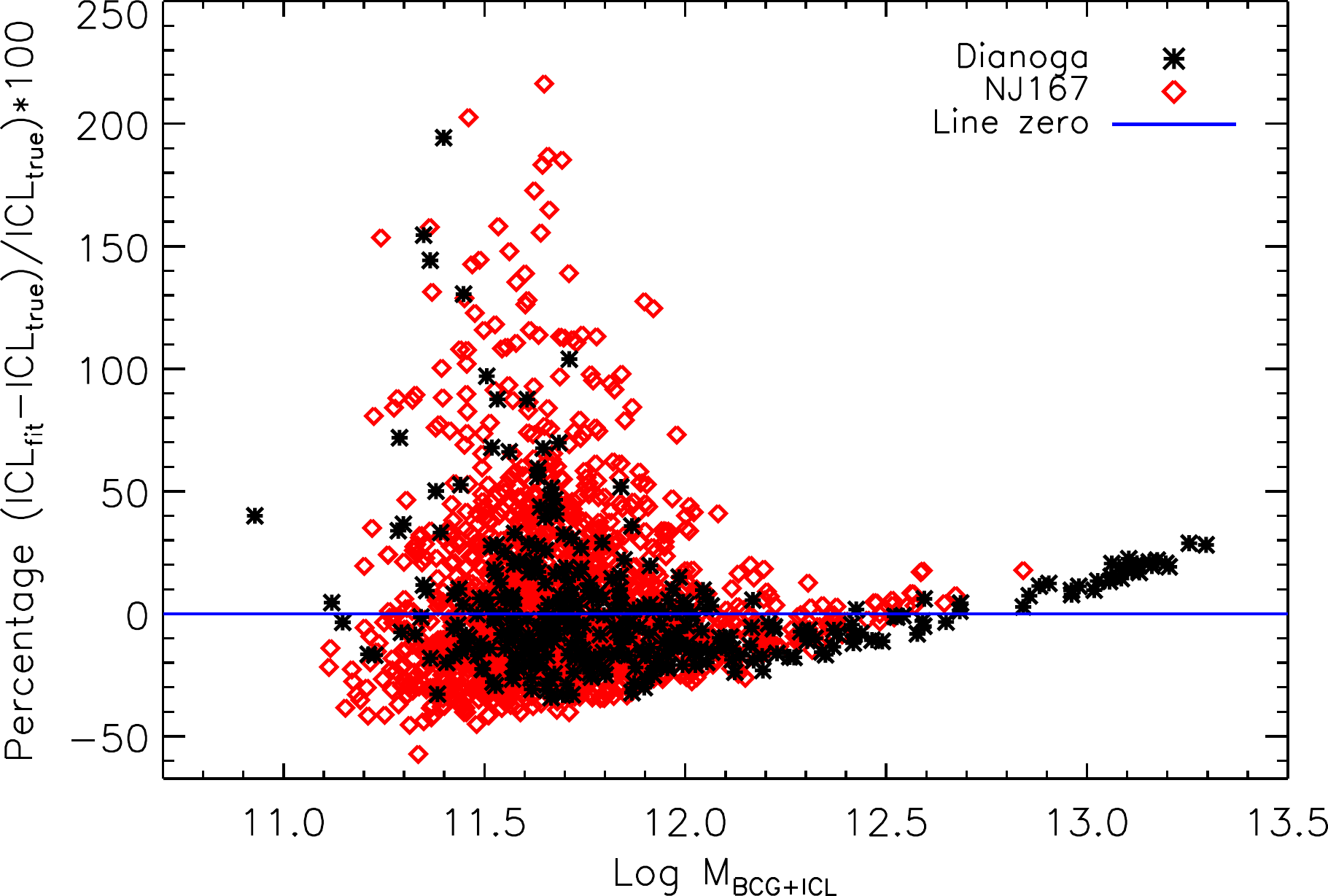} &
\includegraphics[scale=.45]{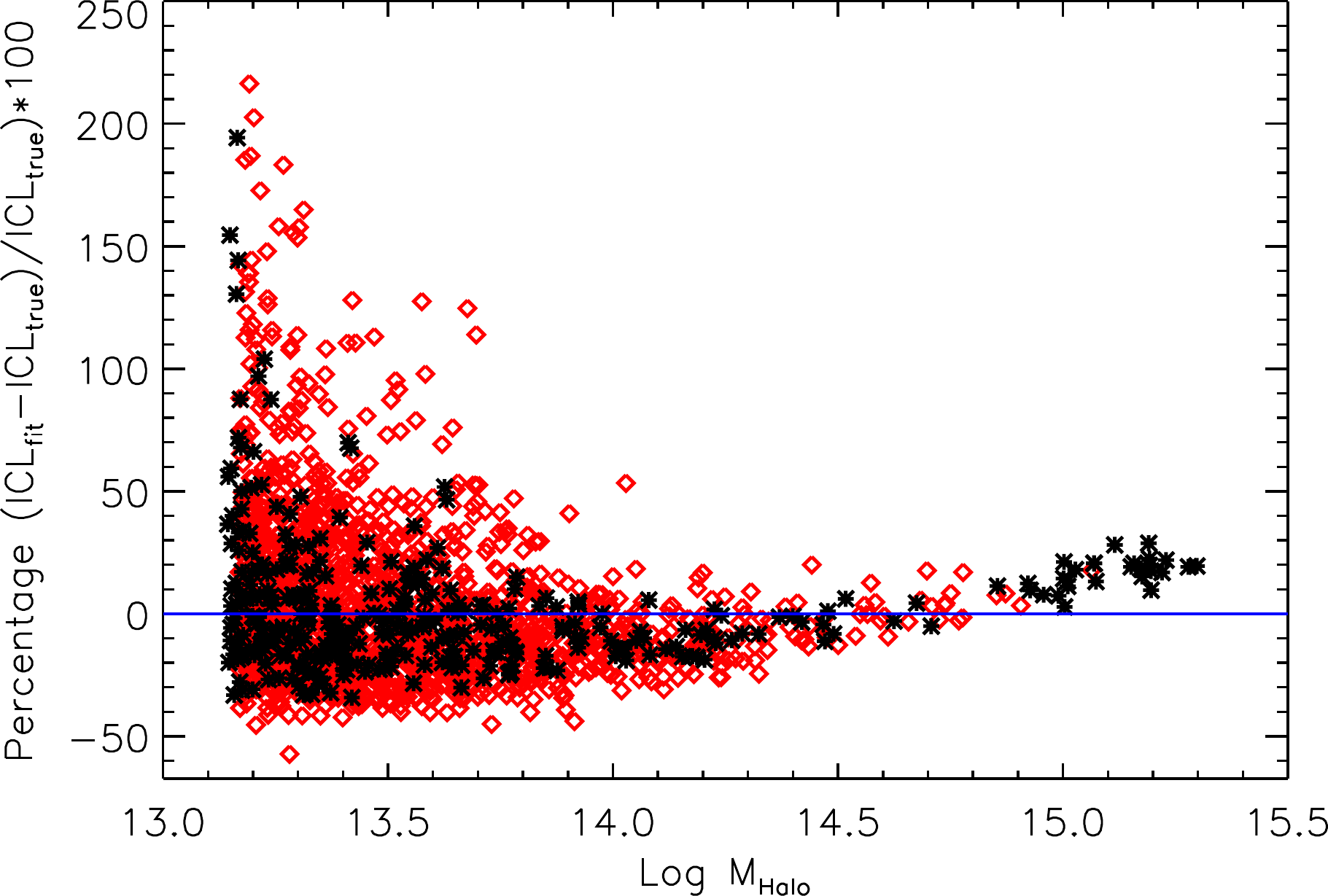} \\
\end{tabular}
\caption{Difference in percentage between the ICL mass derived by using the fits and the true values, as a function of BCG+ICL mass (left panel) and of halo mass (right panel), for Dianoga (black stars) and for 
NJ167 (red squares). The blue lines represent a zero increment along BCG+ICL or halo mass. While the scatter for low mass BCG+ICL ($\log M_{BCG+ICL}<12$) and for halo mass on group scale ($\log M_{Halo}<14$) is 
quite important, on larger BCG+ICL/halo masses it is limited between $\pm$30\%.}
\label{fig:percicl}
\end{center}
\end{figure*}

The aim of the analysis in this section is to find an answer (if possible) to the following questions: assuming we measure the total BCG+ICL stellar mass, is there a way to separate the ICL from the BCG in a 
statistical manner? If yes, what would the typical error be as a function of the total BCG+ICL and/or halo masses? The answer to the first question can be found if and only if the ICL mass correlates well with the 
total BCG+ICL mass, while the answer to the second one strongly depends on how significant the correlation would be. Given the large samples of haloes we can count on, and their wide ranges in terms of mass, we can 
address these points in a statistical way.

Figure \ref{fig:micl_mbcgicl} shows the relation between the stellar mass in ICL versus the total BCG+ICL mass, for our two samples, Dianoga (black stars) and NJ167 (red squares). Overplotted the linear 
fits (black and red lines), done separately for the two sets of data and the details of which are reported in Table \ref{tab:tab2}. The two quantities appear to be in a significant linear relation, with an important 
scatter for objects with $\log M_{BCG+ICL}\lesssim 12.0$ \footnote{This scatter is mainly due to the higher number of groups (which host less massive BCG+ICL systems) with respect to clusters (which host more massive BCG+ICL systems) and to the dispersion in the relation between the halo mass and its concentration, with the latter being an important parameter of the ICL profile.}, but with small scatters in the fit (see the values in Table \ref{tab:tab2}). The two fits are somewhat different, probably due to the different number of haloes and the intrinsic different cosmology of the two simulations, but the slopes of the fits are consistent to each other within $3\sigma$ (and so are the intercepts). 

In order to answer to the questions above, we now want to compare the true value of the ICL mass with that extrapolated from the fit, by knowing just the total BCG+ICL mass. This is done in Figure \ref{fig:micl_fit-true}, where we use the relations reported in Table \ref{tab:tab2} to infer the values of ICL$_{fit}$ to plot against the true values, ICL$_{true}$. Clearly, the data have to stay as close as possible to the blue line in the plot, which represents a one-to-one relation. Indeed, systems with a true value of ICL mass larger than $\log M_{ICL} \sim 11.6-11.7$ tend to populate very close to the blue line, while for smaller ICL masses the scatter is quite large and with a trend of an overestimation of the true values (the value from the fit larger than the true one).

To quantify how much the inferred mass of the ICL from the fit can be reliable (or not), and the trend of the scatter with both BCG+ICL and halo masses, in Figure \ref{fig:percicl} we plot the increase/decrease, in 
percentage, between the value of the ICL mass inferred from the fit and the true ICL mass. The left panel shows this information for both samples (Dianoga with black stars and NJ167 with red squares) as a function 
of BCG+ICL mass, while in the right panel the information is as a function of halo mass. In both panels, the blue line represents a zero increase along BCG+ICL or halo masses. As expected from the previous figure, at 
low BCG+ICL or halo masses, the scatter is wide (see the comment of Figure \ref{fig:micl_mbcgicl}) and the value of the ICL inferred from the fit can be three (and even more) times higher than the true value. However, the important features in this figure is shown by the right sides of both panels. Indeed, for $\log M_{BCG+ICL}>12$ and for halo masses on cluster scales ($\log M_{Halo}>14$), the scatter is limited between $\pm$30\%. We will come back on this point in the next section with a full discussion of the implications of these results.

\section{Discussion}
\label{sec:discussion} 

The aim of this work is to shed some light, from a theoretical point of view, on two aspects that are related to the separation between the mass in the BCG and that of the ICL. In a rough manner, one way could be by
defining a radius at which the ICL dominates over the BCG. Such a method would underestimate the amount of ICL because the two components, BCG and ICL, overlap in a region that can be more or less extended, but it will provide an idea of how extended the region where the ICL distribution would be. 

Another possibility is given by the predictions of theoretical models of the relation that links the amount of ICL with that of the total, BCG+ICL. Indeed, if such a relation is well defined and characterized by a 
small scatter, in principle it is possible to quantify, more or less precisely depending on the scatter, the stellar mass of the ICL from the observed total mass. Here we discuss both possibilities in the light of 
the results described in the previous section and how they compare with observed measurements.

\subsection{The Transition Radius between BCGs and ICL}
\label{sec:r_trans}

The definition of a transition region from the BCG to the ICL is not as trivial as it could seem. Indeed, the two components overlap in a region that can be quite extended depending on the profiles they follow. The 
observational methods used to isolate the ICL have all pros and cons, as argued in Section \ref{sec:intro}, which might bring to underestimate/overestimate the amount of stellar mass in ICL. A simple cut in surface 
brigthness, for instance, could severely underestimate the ICL in the innermost regions that are BCG dominated and similarly, a profile fitting might give the same problem simply because it cannot be precise in the 
region where the two components overlap. 

Our model provides the mass of the BCG and that of the ICL separately, as well as their distributions, which means that we have all the information required to quantify the transition radius of a given BCG+ICL system. We have studied the relations between the transition radius, $R_{transition}$, with BCG+ICL and halo masses, assuming that $R_{transition}$ is the distance from the centre where the ICL starts to account for 90\% of the total (locally) BCG+ICL mass. We also tried several other percentages and in all cases (see Table \ref{tab:tab1}), the transition radius does not depend on BCG+ICL and halo masses. In the case shown in figures \ref{fig:rtrans_bimass} and \ref{fig:rtrans_halomass}, where the percentage is 90\%, the transition radius has an average value of around 60 kpc and a scatter of about $\pm$40 kpc (the precise values depend on the set of simulations), which means that, on average, it can be as small as 20 kpc and as large as 100 kpc. These values are in good agreement with previous results (discussed below in this section), but given the consistency and the large scatter around the mean value, they also point out the difficulty of finding a proxy that can constrain the region of transition for a given BCG+ICL system. 

The transition radius, as mentioned above, can be rougly estimated also from observational studies via profile fitting, similarly to what our model does. A way that can help in measuring $R_{transition}$ is the use 
of a proxy that is directly observable, such as a radius that contains a given percentage of the total BCG+ICL mass. The advantage of this approach consists on the fact that there would not be any need to separate 
the two components. If this radius, which we called BCG+ICL scale radius, $R_{BCG+ICL}$, depends on the BCG+ICL or halo masses, once overplotted with $R_{transition}$, the region of the overlap is an indication of 
the transition radius for a given BCG+ICL/halo mass range obtained directly without separating the two components. Moreover, given the fact that the scale radius can be defined by using different percentages, if 
changing them provides more overlap with the transition radius, we can probe different BCG+ICL/halo mass ranges. 

This was exactly our phisolophy in the analysis done in Section \ref{sec:trans_scale}, where we plot $R_{BCG+ICL}$ by using different percentages together with the transition radius. In our analysis we used 20\%, 
50\% and 65\% to define the scale radius $R_{BCG+ICL}$, and the general trend we find is that an increasing percentage traces the transition radius of decreasing BCG+ICL/halo masses. The scale radius is found to 
increase with BCG+ICL/halo mass, independently of the percentage chosen. This trend has a quite simple explanation: first of all, the mass in a small halo is distributed in a smaller region with respect to larger 
haloes and, secondly, less massive BCG+ICL systems are likely to be hosted by group size haloes, which are also more concentrated than cluster size haloes. A higher concentration of the halo, according to our ICL 
profile (see Equation \ref{eqn:nfw}), reflects in a more concentrated ICL, which in turns means a smaller scale radius. 

Quantitatively speaking, the scale radius $R_{BCG+ICL}$ defined by using different percentages cannot be used as a tracer of the transition radius $R_{transition}$ for different BCG+ICL/halo masses, given the large 
scatters in the relations. Nevertheless, our findings suggest that the region of transition between the BCG and the ICL does not depend on the total mass or halo mass, being universal independently of the system 
considered. This is an important result, considering the fact that an average of 60 kpc can be considered a very small shell within which the BCG dominates the distribution. That region is even smaller if we define
the ICL to dominate the distribution by considering smaller percentages. For instance, if we define $R_{transition}$ as the radius where the local contribution of the ICL is just higher than 50\% of the total 
BCG+ICL mass, $R_{transition}$ becomes, on average, around 20-25 kpc with a scatter of about 10-15 kpc (see Table \ref{tab:tab2}). 

Unfortunately, for the reasons discussed aboved, $R_{transition}$ cannot be taken blindly as the distance after which most of the ICL is distributed for all the range of BCG+ICL/halo masses considered, because 
the region from the centre to the transition radius contains, proportionally, different percentages of ICL from group-size to cluster-size haloes. Indeed, as we can infer from figures \ref{fig:rtrans_bimass} and 
\ref{fig:rtrans_halomass}, while in less massive systems (either in terms of BCG+ICL or halo masses) $R_{transition}$ is much larger than, for example $R_{BCG+ICL,20-50\%}$, in more massive systems it is much smaller. This means that the formers have much more ICL concentrated in the innermost regions (as we already discussed) than the latters. Put in other words, while on cluster scales the transition radius can be considered as the radius from which most of the total ICL is distributed, on group scales it cannot be, because on these scales a larger fraction (with respect to the total) of ICL distributes in the same region where the mass of the BCG does.

Our average values of the transition radius are in good agreement with several observational results. Among the first measurements, it is surely worth mentioning $R_{transition} \sim 80$ kpc by \cite{zibetti05}. The
authors, by stacking the profiles of around 700 galaxy clusters taken from the SDSS, found a flattening of the profile at that radius, and interpreted it as the transition to an ICL dominated region. Such a value was
later confirmed by other works (\citealt{gonzalez05,seigar07,iodice16} and others) which showed a similar break of the BCG+ICL profile in the range 60-80 kpc. More recently, \cite{zhang19} found similar numbers with 
a sample of 300 galaxy clusters in the redshift range $0.2<z<0.3$. These authors analyzed the BCG+ICL radial profile imposing a triple Sérsic model and concluded that the transition from the BCG to the ICL is 
just outside 100 kpc, arguing that this can be interpreted as a potential cut in order to separate the two components. Their conclusion is in good agreement with the discussion above, where we suggest the transition 
radius as a cut to separate BCG and ICL on cluster scales. Close numbers have been found also by \cite{montes21} and \cite{gonzalez21}, $\sim 75$ kpc and $\sim 70$ kpc, respectively. 

Even more recently, \cite{chen21} analyzed a sample of around 3000 clusters at $0.2<z<0.3$ from the SDSS and decomposed the stellar surface mass profile of BCG+ICL in three components, assuming a de Vaucouleurs 
inner profile (BCG dominated region), a outer profile that follows the DM distribution (ICL dominated), and a transitional component that distributes mostly between 70 kpc and 200 kpc. This transitional 
component, argued by the authors to be necessary in order to fully explain the distribution of the ICL, accounts for around 17\% of the total diffuse mass on scales between 50 kpc and 300 kpc, and peaks at 100 kpc 
with 25\%. They interpreted this region as the transition between the BCG dominated part of the distribution, where the ICL forms from processes that are connected to the growth of the BCG itself, such as mergers 
and stripping in the central regions, and the ICL dominated part of the distribution, where the ICL forms via processes that are not linked to the growth of the BCG, such as stripping of infalling satellites and 
pre-processing within infalling groups, that are rather connected to the richness of the cluster. The extent of what the authors called \emph{sphere of influence} (see their paper for more details) can shed some 
light on the processes responsible for the ICL formation.

\subsection{Separation of the BCG and ICL Masses}
\label{sec:bi_mass}

In Section \ref{sec:r_trans} we have seen that the region of transition between a BCG and an ICL dominated parts of the stellar mass distribution of the two components together can be identified by defining a 
transition radius, which is found to be independent of both BCG+ICL and halo masses. However, despite there is no relation between such a radius and either of the two masses, the amount of ICL (with respect to the total BCG+ICL mass) in the region of overlap of the two components can be very different from halo to halo, due to the mass-concentration relation. This means that, even though the transition radius gives an idea of where the ICL starts to dominate the distribution, it cannot be directly taken as a physical separation between the two components in the whole halo (or BCG+ICL) mass range investigated. 

This conclusion leads to the discussion of the analysis done in Section \ref{sec:icl_bcgicl} where, by means of the information provided by our semi-analytic model and our BCG/ICL profile modelling, we analyzed the 
relation between the ICL mass alone and the total BCG+ICL mass, with the goal to provide a way to separate the two components. Clearly, given the fact that the information comes from a semi-analytic model, there are 
caveats that are worth being discussed, and we will do it below in this section. 

From Figure \ref{fig:micl_mbcgicl} and Table \ref{tab:tab2} we can see that the $M_{ICL}-M_{BCG+ICL}$ relation is very tight, scattered at low $M_{BCG+ICL}$ due to the increased number of objects at those scales compared to more massive systems. Two questions arise naturally: given that the $M_{ICL}-M_{BCG+ICL}$ relation is tight, is it possible to extract the ICL mass from it? And if so, what is the typical error that one would make by doing it? We know the total BCG+ICL mass, which is also an observable, and we extract the ICL mass from the relation found in Figure \ref{fig:micl_mbcgicl}. How does the ICL mass extracted from that relation compare with the true ICL mass? Put in other words, we can test the reliability of the fits by comparing the ICL mass extracted from them and the true one. If the two ICL masses are comparable within a reasonable scatter, the fits are reliable and the procedure can be used to separate observed BCG and ICL masses. 

We have addressed this point in Figure \ref{fig:micl_fit-true}, aiming to find an answer to the questions above. We made use of the fits reported in Table \ref{tab:tab2} to derive the ICL mass extracted from the 
$M_{ICL}-M_{BCG+ICL}$ relation and compared it with the true ICL mass. We found that for systems having a true ICL mass larger than $\log M_{ICL} \sim 11.6-11.7$, the fits provide a reasonable estimate of it, while 
below that value the scatter is quite large and the fits tend to overestimate the true value of the ICL mass. These low values of ICL masses likely belong to systems with low BCG+ICL mass, which mostly reside in 
group size haloes. 

In order to quantify the level of accuracy of the fits in extracting the ICL mass from the total, as a function of BCG+ICL and halo masses, in Figure \ref{fig:percicl} we showed the difference in percentage between 
the ICL mass extracted from the fits and the true one, both as a function of BCG+ICL and halo masses. The result is quite clear: in systems with low BCG+ICL mass (or halo mass) the scatter is very large, and such 
a method can underpredict the true value of the ICL mass by 50\% or, even worse, it can overpredict the true value by even a factor of three in the worst cases. Clearly, for these objects the fits do not provide a 
reasonable estimate of the ICL mass, but, at higher BCG+ICL or halo masses, the picture changes. For BCG+ICL systems with $\log M_{BCG+ICL}>12$, or for halo masses on cluster scales, $\log M_{Halo}>14$, the 
difference in percentage between the extracted and true values is confined between $\pm$30\%. Assuming that the scatters in these plots are not mainly driven by the number statistics, $\pm$30\% can be considered a 
reasonable error for a tentative way to separate the BCG from the ICL, in terms of mass estimate. 

Typical errors on the observed estimates of the ICL mass/luminosity in the literature can be higher than the statistical error quoted above. As mentioned a few times in this paper, the most common observational 
methods to isolate the ICL from all the other contributions, BCG included, are all affected by uncertainties that are more or less significative depending on the method used and the quality of the data. Our  
$M_{ICL}-M_{BCG+ICL}$ relation, at least on cluster scale, offers the opportunity to have an indicative idea of the amount of ICL in objects of a given BCG+ICL or halo masses. It is worth noting that the 
$M_{ICL}-M_{BCG+ICL}$ relation (and so the fits) is independent of the particular profiles used to describe the mass distribution of BCG and ICL, since it accounts for the total ICL mass within the virial radius
$R_{200}$. This means that the prediction of our model can be observationally tested, no matter the profile used to describe the mass distribution of the two components. Nevertheless, there are caveats that must be discussed.

In fact, the validity of the results presented in this work depends on the accuracy of our semi-analytic model in predicting the right amounts of the BCG and ICL masses. Starting from the latter, it has been shown 
in several works, theoretical (e.g., \citealt{contini14,tang18,henden20,canas20,asensio20}) but most importantly observational 
(e.g., \citealt{montes14,edwards16,harris17,alamo17,montes18,demaio18,edwards20,demaio20,spavone20,raj20,ragusa21,furnell21,chen21,yoo21}), that our semi-analytic model provides a good description of the formation and evolution of the ICL. Not just its mass at different redshifts, or the fraction as a function of halo mass, but several other properties that have been confirmed by observational studies. Our model is able to predict when the bulk of the ICL starts to form, its colors and metallicity, and the main mechanisms responsible for its formation and evolution. For what concerns the mass of the BCG, this is a long-standing problem. Semi-analytics models, in general, have always had in the past the problem of overpredicting the low and high mass end of the mass/luminosity function, i.e. including the most massive galaxies such as the BCGs. As time passed by, the problem has been largely alleviated by the inclusion of feedback, and in regards of the discussion here, by the inclusion of the ICL formation itself (see, e.g., \citealt{contini14,contini17} and references therein). In the light of these reasons, we believe that the predictions in this work are reliable and directly comparable with observational measurements.

\section{Conclusions}
\label{sec:conclusions}

In this work we have coupled our semi-analytic model of galaxy formation to the merger trees of two sets of N-body simulations in order to have a large statistical sample of haloes in a wide range of mass. We 
applied the model introduced in \citet{contini20} and further improved in \citet{contini21a} for the BCG and ICL mass distributions, and focused our attention on two particular aspects: (a) to investigate the 
region of transition between the BCG and the ICL, a region where the two components overlap and when the latter starts to dominate the distribution; (b) to analyze the relation between the total BCG+ICL mass 
and the ICL mass alone, in order to extract a fitting formula able to separate the ICL mass by knowing the total BCG+ICL one. Concerning the first part of the analysis, i.e. (a), we defined a transition radius,
$R_{transition}$, and a scale radius (observable) of the BCG+ICL systems. The former is defined as the radius where the local contribution of the ICL with respect to the total is 90\%, while the latter is the radius that contains a given percentage of the total BCG+ICL mass. Considering the analysis done in Section \ref{sec:results} and further discussed in Section \ref{sec:discussion}, our main conclusions are as follows:

\begin{itemize}
 \item in a quantitative way, the scale radius $R_{BCG+ICL}$ is not a good proxy of the transistion radius $R_{transition}$, due to the large scatter around the latter. We find $R_{transition}$ to be independent 
       of both BCG+ICL and halo masses, i.e. the distance from where the ICL starts to dominate the distribution is universal and not dependent on the system considered. The typical transition radius predicted 
       by our model is 60$\pm$40 kpc, and lower radii are found by using lower percentages in its definition. These values are in good agreement with recent observational results of BCG+ICL systems residing in 
       cluster size haloes. However, although $R_{transition}$ provides an idea of where the the distribution is dominated by the ICL, it cannot be considered as a physical separation between BCG and ICL in the 
       whole halo, or BCG+ICL, masses probed in this work;
 \item from the analysis of the relation between the BCG+ICL mass and ICL mass alone, we find that, although the two masses correlate, the scatter on low mass systems is non-negligible. By means of a linear fit 
       in log scale, we performed a procedure to separate the two components, isolating the ICL mass directly from the total BCG+ICL one. In systems with low BCG+ICL and/or halo masses, such procedure underpredicts 
       or overpredicts the true value of the ICL in a relevant way, up to a factor of three in the worst cases. However, for systems with $\log M_{BCG+ICL} >12$ or $\log M_{Halo}>14$, the difference between the 
       mass of the ICL extracted from the fit and the true one is confined between $\pm$30\%, meaning that, on cluster scales, the procedure provides a good test for observational methods that try to isolate the 
       ICL mass from the observed BCG+ICL one.
\end{itemize}

In a forthcoming paper, we aim to use a new semi-analytic model with improved prescriptions of several physical processes, such as star formation and feedback, run on the merger trees of a larger cosmological 
simulation (in order to improve the statistics even on cluster scales) to readdress the $M_{ICL}-M_{BCG+ICL}$ relation. Our main goal will be to investigate on whether or not with a more reliable state-of-art semi-analytic model together with much larger statistics the relation will hold on cluster scales. If so, the procedure discussed in this work can be a way, although with non-negligible uncertainties, to separate 
the ICL mass from the observed BCG+ICL mass at least on cluster scales.

\section*{Acknowledgements}
The authors thank the anonymous referee for his/her constructive comments, which helped to improve the manuscript.
EC and QSG are supported by the National Key Research and Development Program of China (No. 2017YFA0402703), the National Natural Science Foundation of China 
(No. 11733002, 12121003, 12192220, 12192222) and  the science research grants from the China Manned Space Project with NO. CMS-CSST-2021-A05.
HZC acknowledge the funds of the cosmology simulation database (CSD) in the National Basic Science Data Center (NBSDC).
\label{lastpage}

\end{document}